\documentclass[aps,prd,floatfix,superscriptaddress,preprintnumbers,nofootinbib,preprint,tightenlines,longbibliography]{revtex4-1}
\pdfoutput=1
\usepackage[english]{babel}
\usepackage{amsmath,amssymb,amsbsy,amstext,amsthm,booktabs,exscale,relsize,slashed,graphicx,amsfonts,upgreek,xcolor}
\usepackage{multirow,dcolumn,bm,enumerate,url,hyperref}
\usepackage{amsmath,amssymb,amsbsy,amstext,amsthm,booktabs,exscale,relsize,slashed,graphicx,amsfonts,upgreek,xcolor,amsmath}
\usepackage{multirow,dcolumn,bm,enumerate,url,hyperref}
\definecolor{lightsabergreen}{rgb}{.14,.64,.14}
\definecolor{lightgreen}{rgb}{.14,.64,.14}
\definecolor{blue}{rgb}{.14,.14,.64}

\hypersetup{
  colorlinks   = true, 
  urlcolor     = lightsabergreen, 
  linkcolor    = lightsabergreen, 
  citecolor   = lightsabergreen 
}

\newcommand{\diff}{\textrm{d}}

\begin{document}
\title{\bf Old Rocks, New Limits: Excavated Ancient Mica Searches For Dark Matter}

\author{Javier F.  Acevedo}
\thanks{{\scriptsize Email}: \href{mailto:17jfa1@queensu.ca}{17jfa1@queensu.ca}; {\scriptsize ORCID}: \href{http://orcid.org/0000-0003-3666-0951}{0000-0003-3666-0951}}
\affiliation{\smaller The Arthur B. McDonald Canadian Astroparticle Physics Research Institute, \protect\\ Department of Physics, Engineering Physics, and Astronomy, \protect\\ Queen's University, Kingston, Ontario, K7L 2S8, Canada}

\author{Joseph Bramante}
\thanks{{\scriptsize Email}: \href{mailto:joseph.bramante@queensu.ca}{joseph.bramante@queensu.ca}; {\scriptsize ORCID}: \href{http://orcid.org/0000-0001-8905-1960}{0000-0001-8905-1960}}
\affiliation{\smaller The Arthur B. McDonald Canadian Astroparticle Physics Research Institute, \protect\\ Department of Physics, Engineering Physics, and Astronomy, \protect\\ Queen's University, Kingston, Ontario, K7L 2S8, Canada}
\affiliation{\smaller Perimeter Institute for Theoretical Physics, Waterloo, ON N2J 2W9, Canada}

\author{Alan Goodman}
\thanks{{\scriptsize Email}: \href{mailto:alan.goodman@queensu.ca}{alan.goodman@queensu.ca}; {\scriptsize ORCID}: \href{http://orcid.org/0000-0001-5289-258X}{0000-0001-5289-258X}}
\affiliation{\smaller The Arthur B. McDonald Canadian Astroparticle Physics Research Institute, \protect\\ Department of Physics, Engineering Physics, and Astronomy, \protect\\ Queen's University, Kingston, Ontario, K7L 2S8, Canada}
\affiliation{\smaller Max Planck Institute for Plasma Physics, \protect\\ Wendelsteinstra{\ss}e 1, 17491 Greifswald, Germany}

\begin{abstract}
    Minerals excavated from the Earth's crust contain gigayear-long astroparticle records, which can be read out using acid etching and microscopy, providing unmatched sensitivity to high mass dark matter. A roughly millimetre size slab of 500 million year old muscovite mica, calibrated and analyzed by Snowden-Ifft et al.~in 1990, revealed no signs of dark matter recoils and placed competitive limits on the nuclear interactions for sub-TeV mass dark matter. A different analysis of larger mica slabs in 1986 by Price and Salamon searched for strongly interacting monopoles. After implementing a detailed treatment of Earth's overburden, we utilize these ancient etched mica data to obtain new bounds on high mass dark matter interactions with nuclei.
\end{abstract}
\maketitle

\section{Introduction}
\label{sec:intro}
There is an abundance of gravitational evidence for the existence of dark matter, including galactic rotation curves, dark matter's lensing of light in galactic clusters, and the influence of dark matter on primordial perturbations measured in the cosmic microwave background and large scale structure. While there are extensive direct searches for dark matter at low-background detectors located deep underground, we have yet to observe dark matter's non-gravitational interactions. Many direct searches for dark matter have endeavoured to detect dark matter particles scattering against nuclei and electrons, constructing extremely large target volumes and filtering out background processes by operating in ultraclean conditions a few kilometres underground. Such experiments have attained exquisite sensitivity to weakly interacting dark matter. On the other hand, these experiments are at present insensitive to dark matter particles with masses $\gtrsim10^{18}$ GeV, since for this mass the number of dark matter particles traveling through a metre-scale detector over the course of a year is order one \cite{Bramante:2018tos,Acevedo:2020avd}. To gain sensitivity to higher masses, detectors must increase either area or exposure time. 

A few decades ago a sequence of papers \cite{Price:1986ky,SnowdenIfft:1995ke} determined that monopoles and dark matter scattering against a nucleus would result in a detectable swathe of damage in old, cleaved slabs of mica etched with hydrofluoric acid. First, a relatively large area ($\sim$metre$^2$) of muscovite mica was etched and inspected with an optical microscope to search for monopoles \cite{Price:1986ky}, which would deposit a large amount of energy through repeated nuclear scattering. A decade later, after extensive calibration using neutron sources, Ref. \cite{SnowdenIfft:1995ke} was able to exclude single dark matter-nuclear interactions based on the absence of short tracks observed in a 80,720 $\mu$m$^2$ slab of 0.5 Gyr-old muscovite mica. This analysis restricted its scope to dark matter masses $\lesssim 10^3$ GeV, and the results were complementary to spin-dependent dark matter searches in 1995. However as we will see here, despite being smaller than presently-running metre-scale underground experiments, the extremely long exposure time of these mica samples render them sensitive to high mass dark matter beyond the reach of modern dark matter detectors. Recently, some new proposals for mineralogical detection of dark matter and neutrinos have also been discussed in \cite{Drukier:2018pdy,Edwards:2018hcf,Baum:2018tfw,Baum:2019fqm,Ebadi:2021cte}. In particular, preliminary bounds on high mass dark matter scattering with nuclei using data from \cite{Price:1986ky} were first obtained in \cite{Jacobs:2014yca}. This work extends results from Refs.~\cite{Price:1986ky,SnowdenIfft:1995ke,Bramante:2018tos} by considering the largest dark matter masses and cross-sections that can be excluded using lack of crystal damage observed in ancient mica. To do so, we will undertake a detailed computation of the energy lost by a dark matter particle as it travels through the Earth's interior before reaching ancient mica, and carefully revisit certain facets of the \cite{Price:1986ky,SnowdenIfft:1995ke} analyses, including a determination of the maximum depth at which the mica could have been situated, the mica orientation, the dark matter entry angle that optimizes detection for low and high cross-sections, and the effect of different dark matter nuclear scattering models. 

The remainder of the paper is organized as follows; In Section \ref{sec:anmica} we present our analysis of overburden material and multiple dark matter scattering off nuclei, adapted for the location, geometry, and composition of ancient mica studied in \cite{Price:1986ky,SnowdenIfft:1995ke}. In Sections~\ref{sec:MicaScat1} and \ref{sec:MicaScat2} we find limits on high mass dark matter's nuclear interactions using mica data from References \cite{SnowdenIfft:1995ke} and \cite{Price:1986ky}, respectively. In Section \ref{sec:conc} we conclude.

\section{Overburden for Ancient Mica Dark Matter Search}
\label{sec:anmica}

In order to obtain bounds on high mass dark matter from the non-observation of crystal damage in ancient minerals, we must first study how much heavy dark matter is slowed by scattering with nuclei as it traverses the Earth's interior, on its way to interacting with the mica. This will determine what amount of the dark matter is moving fast enough through the mica to leave a detectable swathe of molecular damage. First we address the geometry, orientation, and depth of the mica studied in References \cite{Price:1986ky,SnowdenIfft:1995ke}. In Reference \cite{Price:1986ky}, three mica slabs selected from the British Museum, the Smithsonian Institution, and the Stanford University collection with a combined surface area of 1200 cm$^2$ were analyzed. These were reportedly thin slabs and we assume hereafter that their length and width greatly exceeded their depth $L_y=L_z\gg L_x$. Similarly, Reference \cite{SnowdenIfft:1995ke} studied a cleaved mica slab with length and width $L_y=L_z=284$ $\mathrm{ \mu m}$ and identified low-energy nuclear recoils from WIMPs, which it was determined would leave $\sim 5$ $\mathrm {nm}$ length damage tracks in the cleaved and chemically treated mica. In this work, we will assume that in both cases that the mica was buried $d=20$ km underground (this depth is discussed shortly), and comprised of 58\% oxygen, 12\% aluminum, 16\% silicon, and 5\% potassium by number. The remainder of the elements comprising the mica were ignored in \cite{SnowdenIfft:1995ke}, as they were either too light or too scarce to appreciably affect the search for nuclear tracks left by dark matter collisions.

As a consequence of convective forces and tectonic movement of the Earth's crust over an archaeological timescale, the depth at which the mica was buried over its $\sim$5$\times10^8$ years in the Earth may have shifted. Since in our study we will be considering heavy dark matter with large scattering cross-sections, we are concerned with the amount of Earth's crust above the mineral. Thus we are interested in addressing the question, what is the maximum depth the mica could have been buried, given its observed properties?  When analyzing their mica slabs, Refs. \cite{SnowdenIfft:1995ke,Price:1986ky} both determined that the mica samples they selected had a low degree of thermal annealing. Specifically, these studies determined this low rate of annealing using the ratio of tracks left by the alpha decay of $^{238}$U and the alpha decay of $^{212}$Po, which produced a less destructive alpha that would inelastically bind to aluminum and silicon. Reference \cite{SnowdenIfft:1995ke} found that the ratio of these Po versus U alpha tracks dropped to $0.1 \pm 0.02$ from $0.24$ after being annealed at $200$ $^\circ$C for one hour, from which the authors concluded that the minerals were unlikely to have been exposed to temperatures in excess of $200$ $^\circ$C during their habitation underground. 

Using this maximum temperature, $200$ $^\circ$C, we can estimate the maximum depth at which the mica was buried. Below the surface of the Earth, the Earth's crust temperature rises from a temperature $\sim 0^\circ$ C with a gradient $\sim 15-30^\circ$C$/$km \cite{earle2015,JAUPART2015223, arevalo:2009}, although there are some models which predict temperatures as low as $250^\circ$ C at a depth of 20 km \cite{majorowicz2019thermal}. Therefore, to remain conservative in this study (by taking the maximum overburden allowed by observed mica properties), we will assume the mica was buried at a depth of $\sim$20 km. Of course, the mica samples were recovered much nearer to the surface; if a future analysis somehow determines that the mica remained nearer the surface over its $0.5$ Gyr observation time, this could produce a more stringent bound at high dark matter cross-sections. In particular, if we use the less conservative temperature gradient estimate given above to determine a maximum mica depth of $\sim 7$ km, this would result in a factor three improved bound on the maximum overburden cross-section excluded. This is because as can be deduced from (Eq.~\eqref{eq:Echi}), for a fixed experimental threshold energy the upper bound on the DM scattering cross-section will scale linearly with the overburden depth, since both appear in linear proportion in the exponent, which determines number of scatters, $\sigma_{\chi A} \propto L$. So if the depth of the mica was $7$ km, this would correspond to a factor three more restrictive, $i.e.$ larger overburden cross-section.

Having addressed the implications of this annealing temperature for the maximum depth of the mineral, we will make a few comments on the impact annealing may have on the analysis in this document, specifically with regards to whether annealing might erase the recoil damage left by dark matter traversing the mica samples in both Ref.~\cite{Price:1986ky} and ~\cite{SnowdenIfft:1995ke}. In the monopole search Ref.~\cite{Price:1986ky}, the annealing of monopole tracks was investigated using the annealing of ``$\alpha$-interaction'' tracks produced in U and Th decays, where the presence of these tracks determined a threshold energy deposition that could be expected to remain un-annealed. Using the ratio of fission tracks to ``$\alpha$-interaction'' tracks, Ref.~\cite{Price:1986ky} were able to select mica which had no apparent erasure of ``$\alpha$-interaction'' tracks, relative to the number of fission tracks, which left more damage and served as a reliable proxy interaction for the $\alpha$ tracks. Our analysis of heavy dark matter used the \cite{Price:1986ky} energy deposition threshold, which was determined through non-erasure of ``$\alpha$-interaction'' tracks. In the case of the mica WIMP search \cite{SnowdenIfft:1995ke}, the reduction factor for nuclear recoil tracks after thermal annealing at 200 C for 1 hour was found to be $3 \pm 0.5$. Our analysis of \cite{SnowdenIfft:1995ke} explicitly uses the acceptance of nuclear recoil tracks as determined in that analysis, which includes this reduction factor accounting for possible loss of nuclear recoil tracks from thermal annealing.

We now turn to the flux of dark matter through the mica. We are interested in the dark matter flux through the 
widest faces
of the mica, which 
have combined
surface area $A_\oplus=L_yL_z$. We ignore any flux that enters from the other faces of the mica, as they are negligibly small. Taking the local dark matter density to be $\rho_\chi=0.3$ GeV/cm$^3$ \cite{Buch:2018qdr,Lisanti:2016jxe,Iocco:2011jz} and the average velocity of a dark matter particle to be $\langle v_\chi\rangle=320$ km/s \cite{Acevedo:2020gro}, we find the number flux of dark matter over a timescale of $t_\oplus=5\times10^8$ years to be
\begin{align}
    \Phi_\chi=\frac{1}{2}A_\oplus \langle v_\chi \rangle t_\oplus \frac{\rho_\chi}{m_\chi} 
    &\simeq 6.1\times10^{9}\left(\frac{10^{10}\textrm{ GeV}}{m_\chi}\right) \left(\frac{A_\oplus}{80720\textrm{ } \mu\textrm{m}^2}\right) \notag\\
    &\simeq 9.1\times10^{15}\left(\frac{10^{10}\textrm{ GeV}}{m_\chi}\right)\left(\frac{A_\oplus}{1200\textrm{ cm}^2}\right)
    ,
    \label{eq:numFlux}
\end{align}
\noindent for the mica studied in \cite{Price:1986ky,SnowdenIfft:1995ke}. The factor of 1/2 is a geometric acceptance factor which is appropriate for a planar detector.

\begin{figure}[ht]
    \centering
    \includegraphics[width=0.4\linewidth]{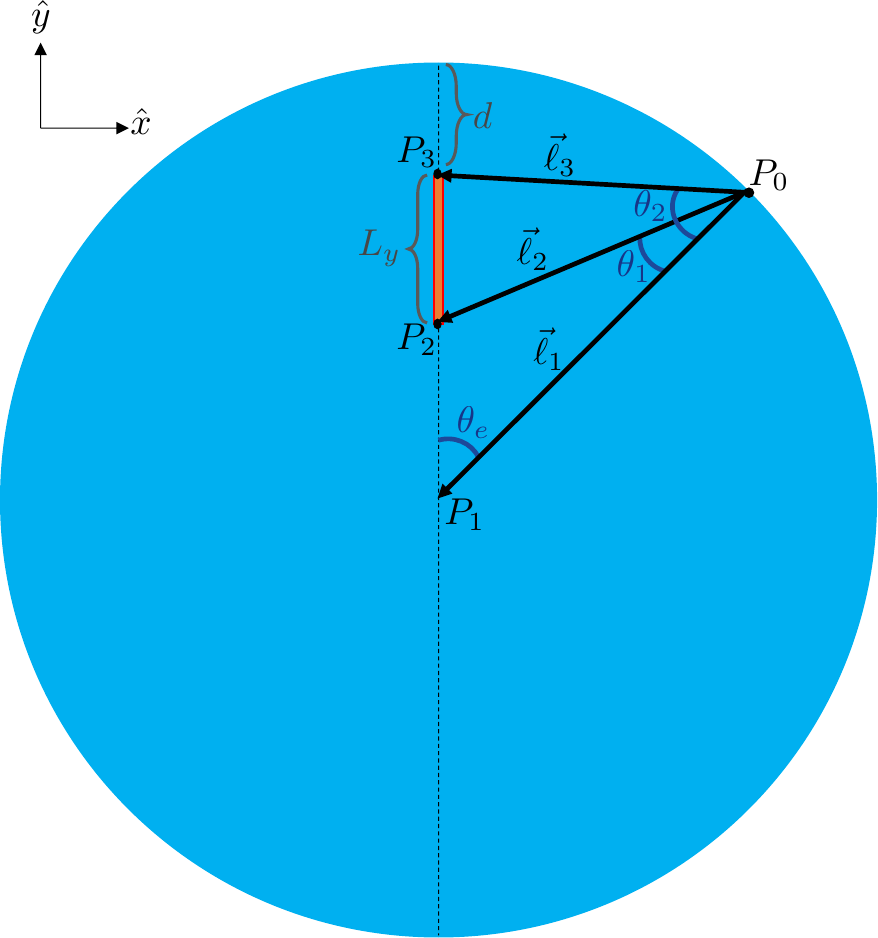}
    \caption{A schematic defining the trajectories for dark matter traversing Earth's overburden in this work, where the blue circle is the Earth and the rectangle is the mica slab. Details and geometric expressions can be found in Appendix \ref{app:Angles}. Diagram not to scale.}
    \label{fig:thetae_schematic}
\end{figure}

Before encountering the mica, each dark matter particle must first travel at least a distance $d$ through the Earth. The actual distance that it travels depends on where the dark matter particle enters the Earth. We will be interested in the probability distribution of lengths travelled by the dark matter before encountering the mica slab. A schematic describing the relevant geometry is given in Figure \ref{fig:thetae_schematic}. Note that in our study we orient the mica such that the vectors normal to its broadest faces point in the direction with the largest overburden, to be conservative, since any other orientation will tend to decrease the effective overburden for incoming dark matter particles. The geometry shown in Figure \ref{fig:thetae_schematic} is detailed in Appendix \ref{app:Angles}.

As it travels to the mica, when a dark matter particle with kinetic energy $E_i$ scatters off of an atom $A$ in the Earth's interior, it will have a final kinetic energy
\begin{equation}
    E_f=E_i\left[1-2\frac{m_\chi m_A}{(m_\chi+m_A)^2} (1-\cos\theta_{cm})\right],
    \label{eq:Ef}
\end{equation}

\noindent where $\theta_{cm}$ is the center of momentum scattering angle, which has a flat probability density. Because the dark matter flux through the Earth is non-relativistic, we simply say that $E_i=m_\chi v_i^2/2$, where $v_i$ is the initial dark matter velocity, which is governed by the distribution given in Ref. \cite{Acevedo:2020gro} as

\begin{align}
    f(v)=\int_{-1}^{1}\textrm{d}\cos\phi\,\frac{\left(v^2-v_e^2\right)^{3/2}}{N_*}\exp\left(-\frac{\tilde{v}^2}{v_0^2}\right)\Theta(v-v_e)\Theta(v_{eg}-\tilde{v}).
    \label{eq:boltzmann}
\end{align}
Here, $\phi$ is the angle between the velocity of the Earth relative to the Milky Way rest frame and the dark matter velocity, $v_0=220$ km/s is the velocity dispersion \cite{Bovy_2012,Read:2014qva,Pato:2015dua}, $v_{eg}=503$ km/s is the escape velocity of the galaxy, and $v_e=11.2$ km/s is the escape velocity of the Earth. Taking $v_{rf}=230$ km/s \cite{Bovy_2012,Read:2014qva} as the velocity of the solar system relative to the galactic rest frame, we define

\begin{align}
    \tilde{v}^2\equiv v^2-v_e^2+v_{rf}^2+2v_{rf}\sqrt{v^2-v_e^2}\cos\phi.
    \label{eq:vtilde}
\end{align}
Note that this distribution has been shifted by $v_e$ to account for dark matter accelerating into the Earth's gravitational potential \cite{LEWIN199687,Acevedo:2020gro}. Finally, we choose $N_*$ numerically such that $\int_{0}^\infty f(v)\,\textrm{d}v=1$. All of these quantities are chosen conservatively at their $1\sigma$ lowered values. We have not accounted for Earth's orbit about the Sun, or its axial rotation, in this distribution, but both of these contributions are negligible. 
When considering a large number of scatterings (say, $\langle\tau\rangle>10^8$), we will not be able to simulate each scattering, and will use an averaged expression for the dark matter's energy loss when scattering in this overburden. Because $\theta_{cm}$ is uniformly distributed, when computing the effect of a large number of scatterings it is accurate to fix $\cos\theta_{cm}\simeq\langle\cos\theta_{cm}\rangle=0$. Hence, the dark matter kinetic energy when reaching the mica will be given by

\begin{equation}
    E_\chi=E_i \prod_A \left[1-2\frac{m_\chi m_A} {(m_\chi+m_A)^2}\right]^{\tau_A}.
    \label{eq:Echi}
\end{equation}

Having defined the mica and Earth geometry, along with the energy loss of dark matter as it travels to the mica, we now turn to the trajectories dark matter will take as it travels to the mica. We want to find the fraction of trajectories that each entry point on the Earth contribute to the total dark matter flux through the mica slab, while accurately accounting for the fraction of dark matter particles that will be undetectable after being slowed by scattering with the intervening Earth material. The relative fraction of flux contributed at a given angle, $\theta_e$, is given by the weighting $f(\theta_e)$,
\begin{equation}
    f(\theta_e) \propto \int_{\theta_1(\theta_e)}^{\theta_2(\theta_e)}\sin\theta'\cos\theta'\, \textrm{d}\theta',
    \label{eq:PthetaE}
\end{equation}

\noindent where this expression approximates the mica slab as a planar disk of diameter $L_y$ for simplicity. This distribution is normalized such that $\int_{0}^{\pi}f(\theta_e')\,\textrm{d}\theta_e'=1$. In our Monte Carlo of dark matter flux, we will randomly select $\theta_e$ values between 0 and $\pi$, and weight the relative flux coming from the resulting trajectories using Eq.~\eqref{eq:PthetaE}. With $\theta_e$ determined, we draw a straight line from $P_0$ to the center of the mica slab, which will make an angle with $\vec{\ell}_1$ which, as defined, is a normal vector to the Earth's surface. This angle is given by $\theta$ as defined in Eq.~\eqref{eq:theta_new} in Appendix \ref{app:Angles}. In defining our trajectories as straight lines, we are making two approximations about the dark matter's trajectory: (1) that it does not change as it scatters off of the elements in the Earth, and (2) it is not appreciably impacted by Earth's gravity. The former is a reasonable assumption, because we are considering models in which a dark matter particle is significantly heavier than any atom in the Earth \cite{Bramante:2018qbc,Bramante:2018tos}. The latter assumption is valid since the Earth's escape velocity is a small fraction of dark matter's halo velocity.

\begin{figure}[ht]
    \centering
    \includegraphics[width=0.67\linewidth]{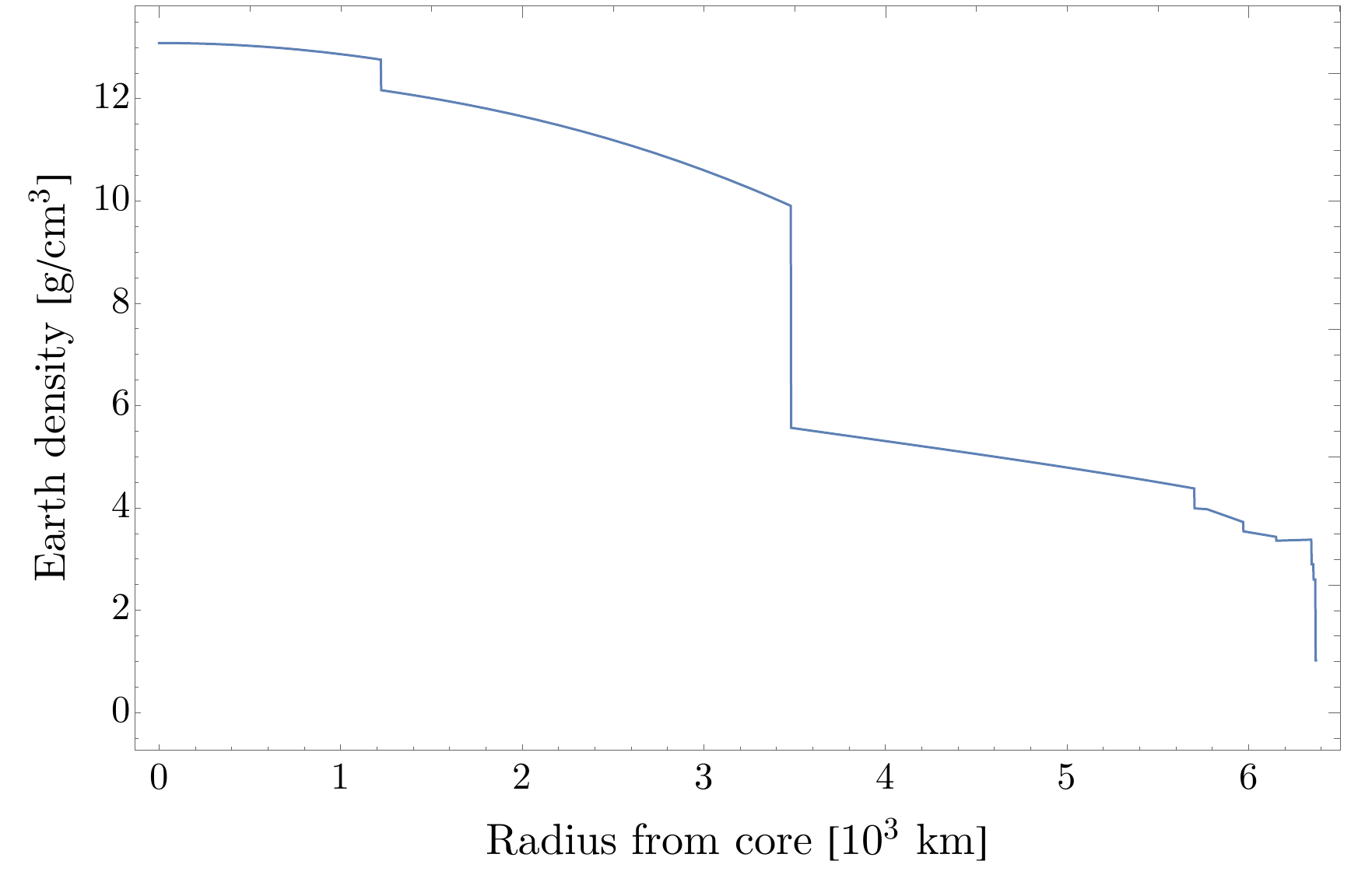}
    \caption{The Earth density used in this work, as a function of radius from the Earth's centre \cite{Bramante:2019fhi,Acevedo:2020gro,Dziewonski:1981xy}.}
    \label{fig:PREM}
\end{figure}

Next we must consider dark matter scattering off a variety of atomic elements in the Earth, which requires us to define the density profile and elemental composition of the Earth. In this work, we use the same model as Refs. \cite{Bramante:2019fhi,Acevedo:2020gro}, which is the Earth density profile in Ref. \cite{Dziewonski:1981xy} with a composition profile given in Refs. \cite{clarke1924composition,WANG2018460,Morgan6973,MCDONOUGH2003547,johnston1974}. Specifically, we use the composition profile given in Table \ref{tab:EarthComp}, and the density profile shown in Fig. \ref{fig:PREM}.

\begin{table}[ht]
\begin{center}
\caption{Rounded weight percentages of elements of interest in the crust, mantle, and core \cite{clarke1924composition,Morgan6973,WANG2018460,MCDONOUGH2003547,johnston1974}. We approximate Earth's crust as a monolithic shell extending from $r=R_\oplus=6371$ to $r=6346$ km \cite{clarke1924composition}. Beneath the crust the mantle extends to $r=3480$ km \cite{WANG2018460}, below this is the core \cite{Morgan6973}.} 
\label{tab:EarthComp}
\resizebox{\textwidth}{!}{
\begin{tabular}{c c c c c c c c c c c c c c}\hline
& \textbf{$^{16}$O} & \textbf{$^{28}$Si} & \textbf{$^{27}$Al} & \textbf{$^{56}$Fe} & \textbf{$^{40}$Ca} & \textbf{$^{23}$Na} & \textbf{$^{39}$K} & \textbf{$^{24}$Mg} & \textbf{$^{48}$Ti} & \textbf{$^{57}$Ni} & \textbf{$^{59}$Co} & \textbf{$^{31}$P} & \textbf{$^{32}$S} \\ \hline

\hline
\textbf{Crust wt\%} & 46.7 & 27.7 & 8.1 & 5.1 & 3.7 & 2.8 & 2.6 & 2.1 & 0.6 & - & - & - & - \\

\hline
\textbf{Mantle wt\%} & 44.3 & 21.3 & 2.3 & 6.3 & 2.5 & - & - & 22.3 & - & 0.2 & - & - & - \\

\hline
\textbf{Core wt\%} & - & - & - & 84.5 & - & - & - & - & - & 5.6 & 0.3 & 0.6 & 9.0 \\
\hline

\end{tabular}}
\end{center}
\end{table}

The mean number of scatterings against a given element $A$ along this particle's trajectory is given by

\begin{equation}
    \langle\tau_A\rangle = \sigma_{\chi A}\int_0^{L}
    n_A(r)\, \textrm{d}l,
    \label{eq:tauAv1}
\end{equation}

\noindent where $n_A(r)$ is the number density of atom $A$ at a distance $r=\sqrt{l^2+R_\oplus^2-2lR_\oplus\cos\theta}$ away from the Earth's center, $\sigma_{\chi A}$ is the DM-nucleus cross-section, and 

\begin{equation}
    L=\frac{R_\oplus\sin\theta_e}{\sin(\pi-\theta-\theta_e)}
    \label{eq:L}
\end{equation}

\noindent is the length from $P_0$ to the mica slab along this trajectory. Because scatterings are independent and  discrete, they are governed by Poisson statistics, meaning that the probability of scattering exactly $\tau_A$ times against element $A$ is given by

\begin{equation}
    P(\tau_A;\langle\tau_A\rangle)=\frac{e^{-\langle\tau_A\rangle} \langle\tau_A\rangle^{\tau_A}}{\tau_A!}.
    \label{eq:tau}
\end{equation}

Finally, the energy loss and nuclear scattering for dark matter traversing the Earth will be model dependent. In general, the sort of dark matter that has a large mass and large cross-section with nuclei, is likely to be dark matter composed of constituents, $i.e.$ a composite dark matter model, see $e.g.$ \cite{Witten:1984rs,Farhi:1984qu,DeRujula:1984axn,Goodman:1984dc,Drukier:1986tm,Nussinov:1985xr,Bagnasco:1993st,Alves:2009nf,Kribs:2009fy,Lee:2013bua,Krnjaic:2014xza,Detmold:2014qqa,Jacobs:2014yca,Wise:2014jva,Wise:2014ola,Hardy:2014mqa,Hardy:2015boa,Gresham:2017zqi,Gresham:2017cvl,Bramante:2018qbc,Gresham:2018anj,Bramante:2018tos,Ibe:2018juk,Grabowska:2018lnd,Coskuner:2018are,Bai:2018dxf,Digman:2019wdm,Bai:2019ogh,Bramante:2019yss,Bhoonah:2020dzs,Clark:2020mna,Acevedo:2020avd}. In this work, we consider three dark matter interactions with nuclei.

\begin{enumerate}
    \item For some spin-independent dark matter models (see $e.g.$ Ref. \cite{Lewin:1995rx}), $\sigma_{\chi A}$ is converted to a DM-nucleon cross-section, $\sigma_{\chi N}^{(SI)}$. For the heavy dark matter we consider here ($m_\chi \gg m_A$), this conversion is given by

\begin{equation}
    \sigma_{\chi A}\simeq A^4\sigma_{\chi N}^{(SI)},
    \label{eq:sigmaChiN}
\end{equation}
note here that in the regime we consider, nuclear form factors are set to unity.

    \item For large composite dark matter models (see $e.g.$ Refs. \cite{Jacobs:2014yca,Digman:2019wdm,Cappiello:2020lbk,Bhoonah:2020fys,Acevedo:2020gro,Acevedo:2020avd}), the nuclear cross-section is constant for all nuclei, $\sigma_{\chi A}=\sigma_c$. This is appropriate if dark matter is a composite state larger than a nucleus, which scatters elastically with all nuclei it intercepts. In particular, we have recently identified a new class of composite dark matter models very large masses and physical sizes, including a detailed cosmology for their formation in \cite{Acevedo:2020avd}. These large composites are coupled to nucleons through the same light mediator that binds their constituents together, and for large enough couplings would scatter elastically with all incoming nuclei.
    
    \item Finally, dark matter could have a spin-dependent cross-section with protons and/or neutrons \cite{Lewin:1995rx}, although in this work we consider proton and neutron cross-sections ($\sigma_{\chi p}^{(SD)}$ and $\sigma_{\chi n}^{(SD)}$ respectively) separately. In this case, for heavy dark matter, the conversion becomes

\begin{align}
    \sigma_{\chi A}\simeq A^2\frac{4(J_A+1)}{3J_A}\left[a_p \langle S_p \rangle_A\right]^2 \sigma_{\chi p}^{(SD)}\,\,\,\,\,\,\,
        \text{(DM-proton SD cross-sections)}
    \label{eq:sigmaChiNSDp}
\end{align}

\begin{align}
    \sigma_{\chi A}\simeq A^2\frac{4(J_A+1)}{3J_A}\left[a_n \langle S_n \rangle_A \right]^2 \sigma_{\chi n}^{(SD)}\,\,\,\,\,\,\,
        \text{(DM-neutron SD cross-sections)}
    \label{eq:sigmaChiNSDn}
\end{align}

Here $J_A$ is the nuclear spin of atom $A$, $\langle S_p \rangle_A$ and $\langle S_n \rangle_A$ are its average proton and neutron spins, and $a_p = a_n = 1$ being proton and neutron coupling constants respectively. Values of $J$, $\langle S_p \rangle$, and $\langle S_n \rangle$ used in this work, as well as relative abundances of isotopes of elements in Table \ref{tab:EarthComp} for which $J\neq0$, are given in Table \ref{tab:SDcomp}.
\end{enumerate}

\begin{table}[ht]
\begin{center}
\caption{Percent (by number) of isotopes with substantial spin for each element of interest, as well as their nuclear spins, $J$, and their average proton and neutron spins, $\langle S_p \rangle$ and $\langle S_n \rangle$ \cite{Bednyakov:2004xq,Bramante:2019fhi}. }
\label{tab:SDcomp}
\resizebox{\textwidth}{!}{
\begin{tabular}{c c c c c c c c c c c c c c c c}\hline
& \textbf{$^{17}$O} & \textbf{$^{29}$Si} & \textbf{$^{27}$Al} & \textbf{$^{57}$Fe} & \textbf{$^{43}$Ca} & \textbf{$^{23}$Na} & \textbf{$^{39}$K} & \textbf{$^{25}$Mg} & \textbf{$^{47}$Ti} & \textbf{$^{49}$Ti} & \textbf{$^{61}$Ni} & \textbf{$^{31}$P} & \textbf{$^{33}$S} \\ \hline

\hline
\textbf{\% (number)} & 0.4 & 4.7 & 100 & 2.12 & 0.135 & 100 & 100 & 10 & 7.44 & 5.41 & 1.14 & 100 & 0.75 \\

\hline

$\boldsymbol{J}$ & 5/2 & 1/2 & 5/2 & 1/2 & 7/2 & 3/2 & 3/2 & 5/2 & 5/2 & 7/2 & 3/2 & 1/2 & 3/2 \\

\hline
$\boldsymbol{\langle{S_p}\rangle}$ & -0.036 & 0.054 & 0.333 & 0 & 0 & 0.2477 & -0.196 & 0.04 & 0 & 0 & 0 & 0.181 & 0 \\

\hline
$\boldsymbol{\langle{S_n}\rangle}$ & 0.508 & 0.204 & 0.043 & 1/2 & 1/2 & 0.0199 & 0.055 & 0.376 & 0.21 & 0.29 & -0.357 & 0.032 & -0.3\\

\hline

\end{tabular}}
\end{center}
\end{table}

\section{High Mass Dark Matter Sensitivity from the Mica WIMP Search}
\label{sec:MicaScat1}

This section derives high mass dark matter nuclear scattering bounds using the ancient mica data in Ref. \cite{SnowdenIfft:1995ke}. In that work, 80,720 $\mu$m$^2$ of mica was cleaved, etched with acid, and then scanned  with an atomic force microscope. This study used low energy proton and neutron sources for calibration, and was able to differentiate nuclear recoil events from backgrounds coming from radioisotopic nuclear decays, using the summed etch depth of sites exhibiting mica crystal damage. Specifically, the track etch model for dark matter developed in \cite{SNOWDENIFFT1995247} determined through extensive calibration that $\sim$keV/amu nuclear recoils produced damage sites in the cleaved mica slab with etched pit depths $\sim 5$ nm, whereas nuclear decay events produced deeper etched pits. Reference \cite{SnowdenIfft:1995ke} found no signal events in the $4- 6.4$ nm etched pit signal region, and with this result set limits on dark matter scattering with nuclei at relatively low nuclear cross-sections and for dark matter masses around $\sim 100$ GeV.

Here we will consider the case of heavy dark matter at very high cross-sections and masses and find an estimate of 90\% exclusion regions using data from \cite{SnowdenIfft:1995ke}. First it should be emphasized that the region of mica probed for nuclear recoils using chemical etching and an atomic microscope in \cite{SnowdenIfft:1995ke} was not limited to the region of the etched pits, where the etching extended to depths $\sim$30 nm. In fact, the etched detection region was sensitive to any nuclear recoils occurring in the mica around the etched region, where nuclei (oxygen, silicon, etc.) could be scattered by dark matter, after which nuclei could re-scatter with other nuclei, eventually leading to detectable mica damage in the etched detection region. In order to determine what fraction of DM-induced nuclear recoils would result in a detectable track in the mica, Ref.~\cite{SnowdenIfft:1995ke} used a detailed calibration, including simulations, which were validated experimentally using low energy neutron recoils \cite{SNOWDENIFFT1995247}. Here we will not attempt to repeat the calibration undertaken in \cite{SnowdenIfft:1995ke,SNOWDENIFFT1995247}. Instead, we will explicitly use the cross-section bound on each nucleus that \cite{SnowdenIfft:1995ke} obtained, to determine at a fixed cross-section, how many of each kind of nucleus dark matter could scatter with as it traversed the mica, in a manner detectable by the etching procedure. This will be sufficient to estimate conservative exclusion regions for the case that dark matter scatters multiple times while transiting the mica, at high cross-section and high dark matter masses. Mica-transiting dark matter will scatter off of a nucleus $A$ an average of
\begin{equation}
    \langle\tau_A\rangle=n_A\sigma_{\chi A} L_s
    \label{eq:tauAv22}
\end{equation}

\noindent times, and each scatter will result in energy being deposited on a nucleus according to

\begin{equation}
    E_{dep}=2E_\chi\frac{m_\chi m_A}{(m_\chi+m_A)^2} (1-\cos\theta_{cm}).
    \label{eq:Edep}
\end{equation}
When a single dark matter particle traverses the mica slab, it will undergo $\sim \tau$ scatterings, with a Poisson distribution defined by Eq.~\eqref{eq:tau} and \eqref{eq:tauAv22}. 

Now that our dark matter models, Earth geometry, and scattering lengths are defined, we are ready to detail the procedure used to estimate 90\% confidence exclusions on high mass dark matter scattering. First, for every nuclear scattering cross-section bound given in \cite{SnowdenIfft:1995ke}, we determined an effective ``nuclear interaction length," $L_s^{(A)}$, which is the effective length over which dark matter transiting the mica will single-scatter with nuclei in a manner detectable by the mica etching procedure. We determined this length by matching the single-scatter cross-section bound in \cite{SnowdenIfft:1995ke} to the flux of dark matter expected through the mica. Specifically, we 
use Eq.~\eqref{eq:numFlux} and \eqref{eq:tauAv22} and 
solve for $L_s^{(A)}$ such that there is a single-scatter dark matter signal event in 90\% of our simulations of the flux of dark matter through the mica, using the 90\% confidence excluded single-scatter nuclear scattering cross-section indicated in \cite{SnowdenIfft:1995ke}. The effective nuclear interaction lengths we find are $L_s^{(A)} \approx (1.4,5.6,5.6,9.1)\times 10^{-10}~{\rm cm}$ for nuclear interactions with (O, Al, Si, K). This effective interaction length is a few orders of magnitude smaller than length of the etching region, as a result of intrinsic detection efficiency of the etching procedure, as detailed in the calibration paper \cite{SNOWDENIFFT1995247}, along with the annealing efficiency reduction factor of $3 \pm 0.5$ explained in Section \ref{sec:anmica}. We also briefly comment on the recoil energies of high mass dark matter as compared to the $m_\chi \approx 10-10^3$ GeV dark matter studied in \cite{SnowdenIfft:1995ke}. In fact, the recoil energy imparted to nuclei in a single-scatter interaction will be very nearly the same for $m_\chi=10^3$ GeV and much higher mass dark matter, since in the limit $m_\chi \gg m_A$ Eq.~\eqref{eq:Edep} becomes $E_{dep} \propto m_\chi^2 m_A/(m_\chi +m_A)^2 v_{\chi}^2 \approx m_A v_{\chi}^2 $. So we expect that the fiducial acceptance of  single-scatter nuclear recoils for $m_\chi = 10^3$ GeV dark matter can be accurately extrapolated to much higher dark matter masses.

In further detail, our method for finding dark matter-nuclear cross-section exclusions at high utilizes a Monte Carlo sampling of dark matter trajectories through the mica and Earth overburden. Using one of the dark matter cross-section models listed in Eqs.~\eqref{eq:sigmaChiN}-\eqref{eq:sigmaChiNSDn}, we fix the dark matter cross-section, and using Eq.~\eqref{eq:tau}, we sample events to find whether a candidate dark matter traversal scatters against some nucleus $A$ as it moves through the mica. In the event of a dark matter recoil over the $L_s^{(A)}$ effective traversal length, we randomly choose $\theta_{cm}$ and find the corresponding dark matter particle kinetic energy $E_\chi$ after the dark matter has moved through the Earth overburden, using the process described in Sec. \ref{sec:anmica}. We then obtain the energy deposited into the nucleus as a result of this scatter using Eq.~\eqref{eq:Edep}. If this energy exceeds a threshold energy, $E_{th}^{(A)} = 100$ eV, then this particle has deposited enough energy to leave a track in the mica\footnote{In point of fact, 100 eV is probably not enough energy to leave a detectable defect in the mica, given calibration performed in \cite{SNOWDENIFFT1995247}. However, the aim of this analysis is to produce a conservative estimate of the high cross-section limit of the bounds obtained in \cite{SnowdenIfft:1995ke}. We have purposefully chosen a low threshold because our analysis effectively vetoes dark matter which scatters multiple times in the detector above threshold, since these particles could have left more track etch damage in the mica than was predicted for single-scatter dark matter in \cite{SnowdenIfft:1995ke}.}. We use 100 eV as  the lowest possible threshold energy for scattering with mica nuclei, since this is the binding energy of silicon dioxide \cite{2012arXiv1210.0038T}. However, we have verified that varying this threshold between $10$ eV and 1 keV, does not materially affect the high mass, high cross-section bound dark matter bound - in practice the cross-section bound varies by less than 10\% with an order-of-magnitude variation in threshold around 100 eV.
Using this method, we can determine what fraction of incident heavy dark matter particles would have left a nuclear recoil track in the mica. More specifically, after sampling enough events, we can determine for what cross-section, detectable track would have been left in the \cite{SnowdenIfft:1995ke} mica data more than 90\% of the time. However when setting bounds, we must pay attention to the fact that if the dark matter scattered more than once and left damage resulting in a $\gtrsim 6$ nm etched pit, this would not have been detected by \cite{SnowdenIfft:1995ke}, which only searched for single-scatter dark matter events. 
From the preceding discussion is it clear that the bound on high mass dark matter depends on a number of probabilities, listed in Table \ref{tab:Probs}. 

\begin{table}[ht]
\begin{center}
\caption{A list of definitions of the probabilities used to set the lower mass exclusion in this work.} 
\label{tab:Probs}

\begin{tabular}{c | c}

$P_{ND}$ & Probability that a given dark matter model is \textit{not} detected (in this work set to 0.1) \\

\hline

$P_E^{(A)}$ & Probability that a single scatter deposits $E>E_{th}^{(A)}$ into atom $A$ \\

\hline

$P_{\chi A}^{(n)}$ & Probability that dark matter scatters $n$ times with element $A$ \\



\end{tabular}
\end{center}
\end{table}

We note that in this high mass analysis, we are assuming that if dark matter scatters multiple times within the mica, it suffers negligible kinetic energy loss. This is a reasonable assumption because the fraction of the dark matter's initial kinetic energy lost in each scatter is $\sim m_\chi/ m_A$, so after at most a few scatters along the mica detection length $L_x$, its kinetic energy should not change appreciably.
With our probabilities laid out, we can explicitly define the probability that traversing heavy dark matter particles are not detected,

\begin{equation}
    P_{ND}^{{\rm low~ mass}}=\left[1- \sum_A P_E^{(A)}  \left(\sum_{n=1}^{\infty} P_{\chi A}^{(n)} \right)\right]^{\Phi_\chi},
    \label{eq:PNT}
\end{equation}

\noindent where Eq.~\eqref{eq:numFlux} tells us $\Phi_\chi$, and we will set $P_{ND}=0.1$ for a 90\% confidence exclusion. We truncate the second sum in this equation at $n=10^3$ -- the regions we analyze do not predict more than this number of scatters in the mica detection region. $P_{\chi A}$ is obtained using Eqs.~\eqref{eq:tau} and \eqref{eq:tauAv22} with the effective scattering length $L_x^{(A)}$:
\begin{equation}
    P_{\chi A}^{(n)}=\frac{e^{-\langle\tau_A \rangle} \langle\tau_A \rangle^n}{n!}.
    \label{eq:Pchim}
\end{equation}
We note that while there is a detailed nuclear track etch model presented in \cite{SNOWDENIFFT1995247}, which predicts the probability that a recoiling nucleus creates an etchable track when moving through mica. For consistency with our geometric setup and overburden considerations, we do not use this model here, since this model from \cite{SNOWDENIFFT1995247} assumed a certain detector geometry around the mica and would need to be validated with their simulation protocol. In particular, future work will need to perform calibrations like those in \cite{SNOWDENIFFT1995247}, but for multiple nuclear recoils in mica, in close proximity, since the the amount of track etch damage left by recoiling nuclei might not simply scale linearly.  Here instead, we have provided a high cross-section estimate by determining the probability of dark matter scattering with multiple nuclei in a possibly detectable manner using the effective scattering length parameter $L_s^{(A)}$ discussed above.


It remains to determine the probability that the dark matter has enough energy to exceed the conservatively low mica damage threshold we have set, $P_E^{(A)}$. We calculate this numerically. First we generate a large number ($N=10^4$) of initial dark matter entry angles, $\theta_e$, and initial velocities, $v_i$, distributed according to Eqs.~\eqref{eq:PthetaE} and \eqref{eq:boltzmann}, respectively. Then following the procedure laid out in Section \ref{sec:anmica}, we calculate the corresponding kinetic energies of these dark matter particles when they reach the mica slab as given by Eq.~\eqref{eq:Echi}. For each of these energies, we calculate their prospective deposited energy from Eq.~\eqref{eq:Edep}. The fraction of these energies that is greater than $E_{th}^{(A)}$ defines $P_E^{(A)}$.

At sufficiently large $m_\chi$, we must also account for statistical variation in the expected dark matter flux. Specifically, we must ensure that there is a 90\% chance of a single track scattering event as before, while now accounting for the not-insignificant chance of zero dark matter particles passing through the mica slab as $\Phi_\chi\rightarrow1$. To calculate this limit, we modify Eq.~\eqref{eq:PNT} to account for statistical variation in $\Phi_\chi$, which will also be Poisson distributed. Specifically, by recognizing that $\Phi_\chi$ is actually the average dark matter number flux, we can write the probability of having $\varphi_\chi$ dark matter particles go through the mica over 0.5 Gyr as

\begin{equation}
    P(\varphi_\chi; \Phi_\chi)=\frac{e^{-\Phi_\chi} \Phi_\chi^{\varphi_\chi}}{\varphi_\chi!}.
    \label{eq:probphi}
\end{equation}
With this, $P_{ND}$ becomes

\begin{equation}
    P_{ND}^{{\rm high~ mass}}=\sum_{\varphi_\chi=0}^\infty P(\varphi_\chi; \Phi_\chi) \left[1- \sum_A P_E^{(A)}  \left(\sum_{n=1}^{\infty} P_{\chi A}^{(n)} \right)\right]^{\varphi_\chi}.
    \label{eq:PNT2}
\end{equation}
We truncate this sum at $\varphi_\chi=10\Phi_\chi$, rounded up to the nearest integer. Due to computational limitations, we used Eq.~\eqref{eq:PNT} for $m_\chi<10^{17.5}$ GeV, and Eq.~\eqref{eq:PNT2} for $m_\chi\geq10^{17.5}$ GeV. With these equations, we find the values of $\sigma_{\chi N}$ or $\sigma_{c}$ for which $P_{ND}=0.1$ for each given $m_\chi$, and exclude all parameter space bounded by these values.

\begin{figure}[h!]
    \centering
    \includegraphics[width=\linewidth]{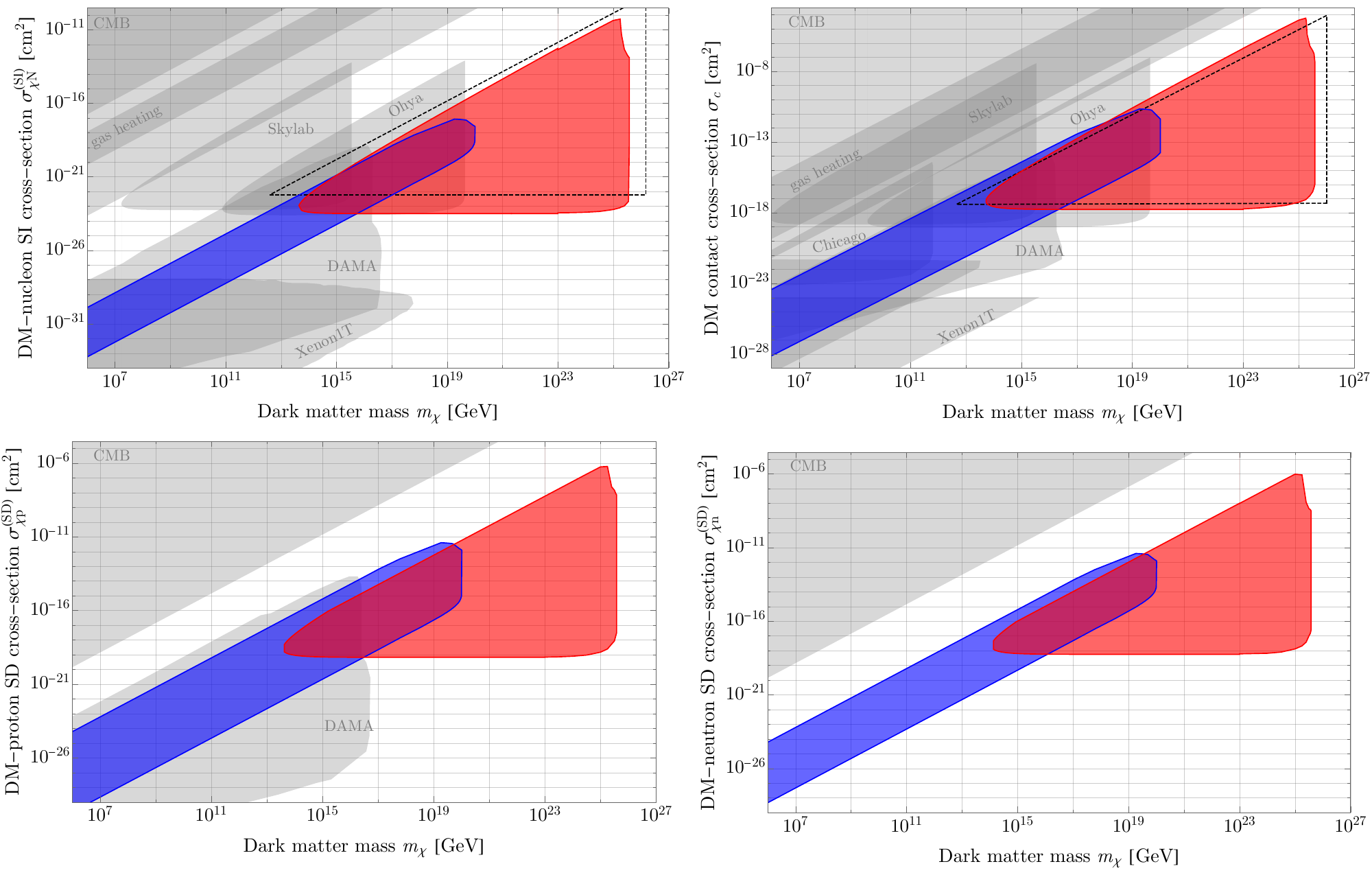}
    \caption{High mass dark matter cross-section exclusions are shown, from new analyses of mica monopole data \cite{Price:1986ky} (red) and an etched mica WIMP search \cite{SnowdenIfft:1995ke} (blue). Other relevant exclusions are shown in grey. 
    \textbf{Top Left:} Spin-independent DM-nucleon cross-section exclusions. Juxtaposed limits include those from XENON1T \cite{Clark:2020mna} (see also LUX \cite{Akerib:2016vxi} and PandaX-II \cite{Cui:2017nnn}), DAMA \cite{Bernabei:1999ui}, the CMB \cite{Dvorkin:2013cea, Gluscevic:2017ywp}, interstellar gas cloud measurements \cite{Bhoonah:2018gjb,Bhoonah:2020dzs}, and exclusions from detectors at the Skylab Space Station and the Ohya quarry near Tokyo \cite{Bhoonah:2020fys}. The dashed region is a prior estimate of the monopole mica exclusion \cite{Bramante:2018tos}.
    \textbf{Top Right:} Contact cross-section exclusions. Juxtaposed limits include a new experiment labelled Chicago \cite{Cappiello:2020lbk}, DAMA \cite{Bernabei:1999ui}, detectors at the Skylab Space Station and the Ohya quarry near Tokyo \cite{Bhoonah:2020fys}, and an interstellar gas cloud heating exclusion \cite{Bhoonah:2020dzs}. The latter of these bounds reports re-scaled exclusions (from DM-nucleon to contact exclusions) for XENON1T \cite{Aprile:2017iyp,Aprile:2019dbj}, \cite{Price:1986ky}, and CDMS-I \cite{Abusaidi:2000wg,Abrams:2002nb}. The dashed region is a previous exclusion estimate obtained in Ref. \cite{Jacobs:2014yca,Bhoonah:2020dzs}.
    \textbf{Bottom Left: } The spin-dependent DM-proton cross-section exclusions, alongside DAMA \cite{Bernabei:1999ui} and CMB \cite{Dvorkin:2013cea, Gluscevic:2017ywp} exclusions.
    \textbf{Bottom Right: } The spin-dependent DM-neutron cross-section exclusions, alongside CMB exclusions \cite{Dvorkin:2013cea, Gluscevic:2017ywp}. Additional high mass spin-dependent bounds from underground experiments should be obtainable using, $e.g.$, existing XENON1T and PICO \cite{Amole:2015pla} data.
    }
    \label{fig:Excls}
\end{figure}

The blue limits in Figure \ref{fig:Excls} describe the regions in which $m_\chi>10^6$ GeV and dark matter scattering with the indicated cross-section can be excluded at 90\% confidence level, $i.e.$ $P_{ND}\leq0.1$. We consider a spin-independent nucleon cross-section $\sigma_{\chi N}^{(SI)}$, a contact cross-section $\sigma_{c}$, and spin-dependent proton and neutron cross-sections $\sigma_{\chi p}^{(SD)}$ and $\sigma_{\chi n}^{(SD)}$ respectively. There are four primary factors that determine this exclusion's boundaries:
\begin{enumerate}
    \item The lowest cross-sections that we exclude are limits taken directly from \cite{SnowdenIfft:1995ke}, which gives a 90\% confidence lower-bounds on dark matter scattering with all the major component nuclei of their mica sample. For cross-sections smaller than this, the expected number scattering events is small, and thus the likelihood of a detectable track being created is small as well.
    
    \item Most of the high cross-section limit is set by the maximum cross-section at which the dark matter still has enough energy to leave a track when scattering against an atom in the mica. At larger cross-sections, it will have lost too much energy from scattering in the overburden before reaching the mica (as described in Sec. \ref{sec:anmica}), to leave a detectable track.
    
    \item At the highest dark matter masses and cross-sections (upper right of the plot), the exclusion region is limited by multi-scatter events. If dark matter scatters too much in the $L_x$ readout region, these events would have been discarded as background in the analysis of \cite{SnowdenIfft:1995ke}.
    
    \item The highest dark matter masses in our exclusion region are flux limited. At sufficiently high $m_\chi$, fewer than one dark matter particle is expected to have entered the mica slab in its half gigayear lifetime.

\end{enumerate}

\section{High Mass Dark Matter Sensitivity from the Mica Monopole Search}\label{sec:MicaScat2}

Next, we revisit high mass dark matter exclusions using the mica monopole search laid out in Ref. \cite{Price:1986ky}. Some of the details of this exclusion will be different from those in Section \ref{sec:MicaScat1}, since in the monopole search \cite{Price:1986ky} the energy threshold to create a track in the mica was set not by single scattering events, but by multiple scattering events parameterized by an energy deposition rate. Specifically, the energy deposition threshold, calibrated using nuclear recoil data in \cite{Price:1986ky} is given by 

\begin{equation}
    \frac{\diff E}{\diff x}\bigg|_\textrm{th} \simeq \frac{7\textrm{ GeV/cm}}{\cos\theta_m}.
    \label{eq:dEdxth}
\end{equation}
Here, $\theta_m$ is the zenith angle that a dark matter particle makes with the surface of the mica slab. In terms of the Earth geometry already introduced, $\theta_m=\pi/2-\theta_e-\theta$, where $\theta$ is the angle between a straight line drawn from $P_0$ to the center of the mica slab and $\ell_1$, as detailed in Appendix \ref{app:Angles}, which can be compared to angles shown in \ref{fig:thetae_schematic}. The total rate at which the dark matter with energy $E_\chi$ deposits energy into the mica is \cite{Bhoonah:2020fys}
\begin{align}
    \frac{\diff E}{\diff x}&=\frac{2 E_\chi}{m_\chi}\sum_A\frac{\mu_{\chi A}^2}{m_A} n_A \sigma_{\chi A} \notag\\ &\simeq 1.5\textrm{ GeV/cm}\left(\frac{\sigma_{c}}{10^{-18}\textrm{ cm}^2}\right)\left(\frac{v_\chi}{0.001}\right)^2,
    \label{eq:dEdx}
\end{align}
\noindent where $\mu_{\chi A}=m_\chi m_A/(m_\chi+m_A)$ is the reduced mass of the dark matter and atom $A$. If $\diff E/\diff x>\diff E/\diff x|_\textrm{th}$ for a given dark matter particle, a detection event is said to have occurred. Since zero detection events were observed in the monopole mica search, as in Section \ref{sec:MicaScat1} we exclude dark matter models for which $P_{ND} \leq 0.1$.

\emph{High mass dark matter.} We start by considering dark matter masses high enough to permit explicit simulation of the entire dark matter flux expected through the mica. Specifically, for $\Phi_\chi\ll 10^4$, we can directly simulate these scenarios using less than $10^4$ events for a given dark matter mass and cross-section, generated using the same method laid out in Section \ref{sec:MicaScat1}. As can be verified by inspecting the total integrated flux at the mica monopole search, cf.~Eq.~\eqref{eq:numFlux}, this corresponds to masses $m_\chi\gtrsim10^{23}$ GeV.

For each dark matter particle trajectory simulated (holding mass and cross-section constant), we randomly selected an initial dark matter velocity from Eq.~\eqref{eq:boltzmann}, as well as a random trajectory $\theta_e$ from $10^4$ events simulated as described in Section \ref{sec:MicaScat1}. Then, using Eq.~\eqref{eq:Echi}, we calculated the kinetic energy at the mica slab, to determine whether the particle would have been detected. We sampled a single trial flux by repeating this process $\varphi_\chi$ times, where $\varphi_\chi$ is determined from Eq.~\eqref{eq:numFlux}. After generating 1000 trial fluxes for a given dark matter cross-section and mass, we excluded a mass and cross-section pairing if $>$90\% of the trials generated any detection events at the monopole mica. The red excluded region in Fig.~\ref{fig:Excls} shows the region for which this criteria was met.

\begin{figure}[ht]
    \centering
    \includegraphics[width=\linewidth]{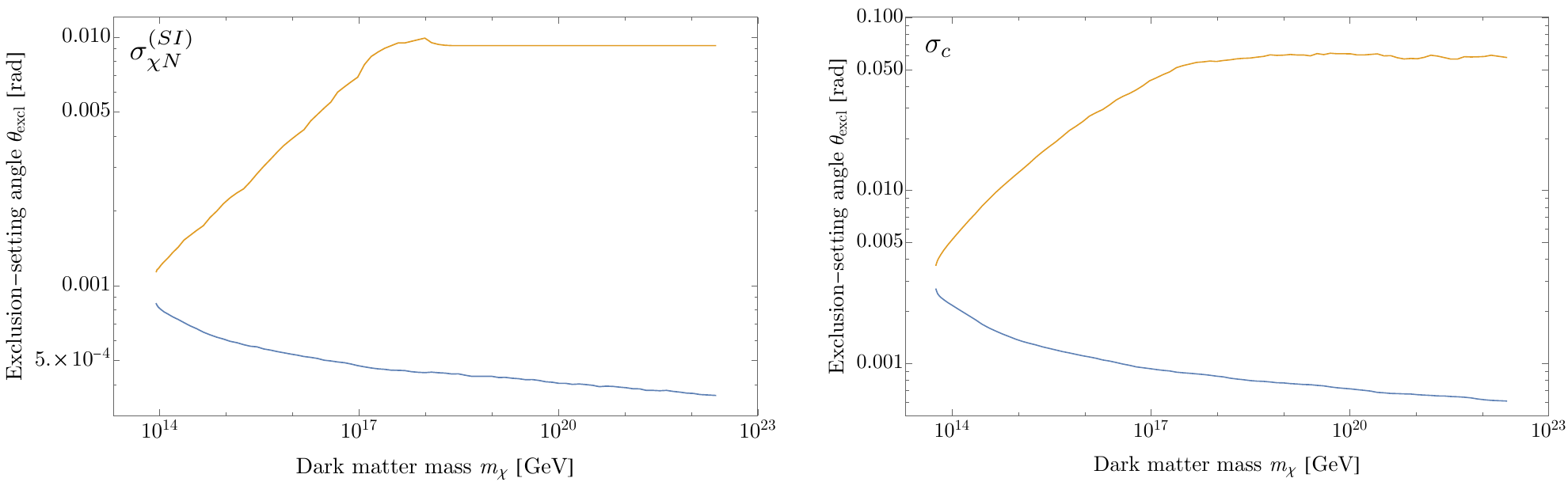}
    \caption{The Earth entry angle along which dark matter most easily exceeds the mica damage threshold, $\theta_\textrm{excl}$ is plotted as a function of $m_\chi$ for the case of dark matter with a contact nuclear cross-section ($\sigma_c$) and a spin-independent nucleon cross-section ($\sigma_{\chi N}^{(SI)}$). The (upper) yellow lines correspond to a low cross-section bound, where longer paths through the overburden permit a more perpendicular impact with the mica. A more perpendicular impact with the mica allows for etching to become evident at a lower threshold scattering energy, Eq.~\eqref{eq:dEdx}. The (lower) blue lines correspond to a high cross-section bound, which requires a shorter path through the Earth overburden. Both lines show moving averages of 10 actual $(m_\chi,\theta_\textrm{excl}$) pairs.
    }
    \label{fig:thetaExcls}
\end{figure}

\emph{Low mass dark matter.} For dark matter masses $m_\chi\leq10^{23}$ GeV it was untenable to simulate the entire dark matter flux explicitly. To obtain a bound in this regime, we used an expression for for the energy deposition rate of dark matter into the mica detector ($D$) after traversing an overburden ($O$) \cite{Bhoonah:2020fys}
\begin{equation}
    \frac{\diff E}{\diff x}=\frac{2 E_i}{m_\chi}\left(\sum_{A\subset D} \frac{\mu_{\chi A}^2}{m_A}n_A\sigma_{\chi A} \right)\exp\left(-\frac{2}{m_\chi}\sum_{A\subset O}\frac{\mu_{\chi A}^2}{m_A}\tau_A\right).
    \label{eq:dEdxFull}
\end{equation}
This formula relies on a single dark matter energy at Earth's surface, $E_i$, and a single dark matter trajectory, parameterized by $\tau_A$, which in turn depends depends on $\theta_e$ in Figure \ref{fig:thetae_schematic}. When $\Phi_\chi$ is relatively large (as it is for lower dark matter masses), the limiting cross-section will be determined by the cross-section for which a small number of dark matter particles travel along optimal trajectories that result in an etched track at the monopole mica search. Specifically, there will be a certain angle $\theta_e$, that minimizes scattering in the overburden, while still allowing for the dark matter to trigger the etched mica search, given that $\theta_e$ also affects the threshold cross-section for creating etchable damage in mica, cf.~Eq.~\eqref{eq:dEdx}.

To set our lower mass exclusions we fixed $m_\chi$, and then scanned over cross-section values $\sigma_\textrm{test}$, comparing $\diff E/\diff x$ to $\diff E/\diff x|_\textrm{th}$ for all values of $\theta_e$. If, for example for all $\theta_e$, $\diff E/\diff x < \diff E/\diff x|_\textrm{th}$, this $\sigma_\textrm{test}$ could not be excluded. On the other hand, if there existed even a very narrow range of $\theta_e$ for which $\diff E/\diff x > \diff E/\diff x|_\textrm{th}$ (but still wide enough such that more than one heavy dark matter particle satisfying this condition would intercept the mica over 500 Myr), then $\sigma_\textrm{test}$ was excluded. The $\sigma_\textrm{test}$ for which $\diff E/\diff x = \diff E/\diff x|_\textrm{th}$ over a small range of $\theta_e$ is the cross-section that defines the border to the excluded region. We call the smaller angle bounding this range $\theta_\textrm{excl}$. To illustrate this method, we show this angle as a function of $m_\chi$ in Figure \ref{fig:thetaExcls}, for spin-independent and contact cross-sections with nuclei. There are two sets of solutions. One set of solutions minimizes overburden scattering at high cross-section, corresponding to small $\theta_e$. The second set of solutions tends towards a wider angle that gives a more perpendicular dark matter impact with the mica, to allow for etchable damage to occur at lower scattering cross-sections (note in Eq.~\eqref{eq:dEdx} that the mica threshold is a function of entry angle). We note that the exclusions set using this method, agree with the Monte Carlo method used for high mass dark matter. Both are shown in Figure \ref{fig:Excls}, where bounds set using the two methods join at $m_\chi=10^{23}$ GeV. The union of these exclusions is shown in red in Figure \ref{fig:Excls}.

\section{Discussion}
\label{sec:conc}
Using ancient mica data laid out in \cite{Price:1986ky,SnowdenIfft:1995ke}, we have obtained new bounds on heavy dark matter for masses up to $10^{25}$ GeV for spin-independent, spin-dependent, and contact interactions. To obtain accurate limits we undertook detailed calculations of Earth's overburden, and recalibrated ancient mica data for use with simplified models of superheavy dark matter interactions with nuclei. Our findings provide some interesting insights into the behaviour of ancient mineral exclusions in multiscatter and flux-limited regimes, especially using data from the mica search for monopoles \cite{Price:1986ky}. Furthermore, we found that the finer-grained ancient etched mica data in \cite{SnowdenIfft:1995ke} can place bounds on rather heavy dark matter that only scatters once inside a thin mica slab.

It is striking that ancient etched mica data collected decades ago still places world-leading bounds on models of large composite dark matter interactions with nuclei. In future, it would be interesting to study how sensitive these mica searches are for models of large composite states that couple very weakly to the Standard Model through a light mediator \cite{Acevedo:2020avd}. This work has particularly shown the utility of minerals with archaeological-scale exposures, as a detection bed for very rare interactions of high mass dark matter. As such, a dedicated effort to use minerals and other long-lived materials as low-background dark matter detectors will be a fruitful and useful endeavour in exploring the high mass, high cross-section, and composite dark matter frontiers.

\section*{Acknowledgements}
We thank Levente Balogh, Matthew Leybourne, Bill McDonough, Aaron Vincent, and an anonymous referee for useful discussions and correspondence. The work of JA, JB, AG is supported by the Natural Sciences and Engineering Research Council of Canada (NSERC). Research at Perimeter Institute is supported in part by the Government of Canada through the Department of Innovation, Science and Economic Development Canada and by the Province of Ontario through the Ministry of Colleges and Universities.

\appendix

\section{Earth and Mica Slab Geometry}
\label{app:Angles}
This appendix details the geometry shown in Figure \ref{fig:thetae_schematic}. With $R_\oplus=6371$ km defined as the radius of the Earth and $\theta_e$ as the polar angle defined clockwise from the $y$-axis about the Earth, $P_0=R_\oplus(\sin\theta_e,\,\cos\theta_e)$ is the point where the dark matter enters the Earth, $P_1=(0,0)$ is the center of the Earth, $P_2=(0,\,R_\oplus-d-L_y)$ is the bottom end of the mica, and $P_3=(0,\,R_\oplus-d)$ is the top end of the mica. We connect $P_0$ to $P_1$, $P_2$, and $P_3$ with $\vec{\ell}_1$, $\vec{\ell}_2$, and $\vec{\ell}_3$ respectively ($i.e.$ $\vec{\ell}_n=P_i-P_0$ for $i \in \{1,2,3\}$). The angles $\theta_1$ and $\theta_2$ defined between $\vec{\ell}_1$ and $\vec{\ell}_2$, $\vec{\ell}_2$ and $\vec{\ell}_3$ respectively are

\begin{equation}
    \theta_1(\theta_e)=\arccos\left[\frac{\vec{\ell}_1\cdot\vec{\ell}_2}{|\vec{\ell}_1||\vec{\ell}_2|}\right]=\arccos\left[ \frac{R_\oplus+(d+L_y-R_\oplus)\cos\theta_e}{\sqrt{(d+L_y-R_\oplus\cos\theta_e)^2+R_\oplus^2\sin^2\theta_e}} \right],
    \label{eq:theta1}
\end{equation}

\begin{equation}
    \theta_2(\theta_e)=\arccos\left[\frac{\vec{\ell}_1\cdot\vec{\ell}_3}{|\vec{\ell}_1||\vec{\ell}_3|}\right]=\arccos\left[ \frac{R_\oplus+(d-R_\oplus)\cos\theta_e}{\sqrt{(d-R_\oplus\cos\theta_e)^2+R_\oplus^2\sin^2\theta_e}} \right].
    \label{eq:theta2}
\end{equation}
The angle between zenith is
\begin{equation}
    \theta=\arcsin\left[ \frac{R_\oplus-d-L_y/2}{\sqrt{R_\oplus^2\sin^2\theta_e+(R_\oplus-d-L_y/2-R_\oplus\cos\theta_e)^2}}\sin\theta_e \right].
    \label{eq:theta_new}
\end{equation}

\bibliographystyle{JHEP.bst}

\bibliography{dmica.bib}

\end{document}